\documentclass[pra,twocolumn]{revtex4}

\usepackage{amsmath,amssymb,mathrsfs}
\usepackage{amsthm}
\usepackage{psfrag}
\usepackage{graphicx}
\usepackage{graphics}
\usepackage{epsfig}
\usepackage{bm}
\usepackage{color}
\usepackage{verbatim,color,ulem}
\usepackage{ulem}

\newcommand{\beq}{\begin{equation}}
\newcommand{\eeq}{\end{equation}}
\newcommand{\beqa}{\begin{eqnarray}}
\newcommand{\eeqa}{\end{eqnarray}}

\newcommand{\be}{\begin{equation}}
\newcommand{\ee}{\end{equation}}

\newcommand{\E}{\mathrm{e}}

\newcommand{\ket}[1]{\left| #1\right\rangle}
\newcommand{\bra}[1]{\left\langle #1\right|}
\newcommand{\ave}[1]{\langle{#1}\rangle}

\begin{document}

\title{Coherent Bose-Einstein condensation with fluctuating density}

\date{\today}

\begin{abstract}
Bose-Einstein condensation in the grand canonical ensemble 
admits a formulation in terms of a phase-density decomposition
of the condensate mode operator $\hat{\psi}_{\bf 0}$. 
In the presence of macroscopic condensate number fluctuations
this representation presents nontrivial
implications. In particular, we show that, for the ideal gas, 
under the assumption of a
well-defined phase and a fluctuating condensate density, the modulus squared of the anomalous
average $\langle \hat{\psi}_{\bf 0}\rangle$ 
can provide only a fraction of the whole condensate density $\rho_{\bf 0}$ and 
for the grand canonical statistics of the ideal Bose gas one
obtains the value $|\langle {\hat \psi}_{\bf 0}\rangle|^2
=(\pi/4) \rho_{\bf 0}$. The remaining
part is supplemented by the (macroscopic) fluctuations of 
$\hat{\psi}_{\bf 0}$, which become a distinctive feature of the BEC in this setting. 
This provides a transparent physical picture
of a condensate of photons with a well-defined phase but large 
number fluctuations, as observed in dye-filled microcavity photon experiments.
We also propose a way to access the
square modulus of the anomalous average to test theoretical predictions.
\end{abstract}

\author{L. Salasnich}
\email{luca.salasnich@unipd.it}
\affiliation{Dipartimento di Fisica e Astronomia ``Galileo Galilei'' and
Padua QTech}
\affiliation{Università di Padova, via Marzolo 8, 35131 Padova, Italy}
\affiliation{INFN, Sezione di Padova, via Marzolo 8, 35131 Padova, Italy}

\author{A. Crisanti}
\email{andrea.crisanti@uniroma1.it}
\affiliation{Dipartimento di Fisica, Universit\`a di Roma Sapienza, P.le Aldo Moro 2, 00185 Rome, Italy}
\affiliation{Istituto dei Sistemi Complessi - CNR, P.le Aldo Moro 2, 00185 Rome, Italy}

\author{A. Sarracino}
\email{alessandro.sarracino@unicampania.it}
\affiliation{Dipartimento di Ingegneria, Universit\`a della Campania ``Luigi Vanvitelli'', via Roma 29, 81031 Aversa (CE), Italy}

\author{M. Zannetti}
\email{mrc.zannetti@gmail.com}
\affiliation{Dipartimento di Fisica ``E. R. Caianiello'',
Universit\`a di Salerno, via Giovanni Paolo II 132, 84084 Fisciano (SA), Italy}

\maketitle

\section{Introduction}

The theory of Bose-Einstein condensation has a long history, with foundational contributions by
Bogoliubov, Penrose and Onsager, and Yang
\cite{bogoliubov,penrose,onsager,yang}.
In the standard approach, the condensate is described by a
macroscopically occupied mode with a fixed phase, often
implemented through the Bogoliubov quasi-average construction
\cite{bogoliubov,ginibre}. 

Beyond the average condensate fraction, the study of number fluctuations of the condensate has revealed 
interesting non-trivial features \cite{kruk}, even in the simplest setting of the ideal gas, 
where different thermodynamic ensembles can predict different behaviors \cite{ziff,weiss}. 
In particular, in the grand canonical ensemble (GCE) fluctuations are macroscopic 
and do not go to zero with the temperature, leading to the so-called grand canonical catastrophe \cite{kruk}. 
The interest in the condensate fluctuations has been further enhanced 
by recent progresses in experimental techniques which
allowed to measure fluctuations in different systems \cite{kristensen,christensen,jalali}. 

In particular, a remarkable experiment where the grand canonical statistics of the condensate has been realized, 
leading to strong intensity fluctuations of the ground mode, is the Bose-Einstein condensation of photons in optical dye-filled microcavities \cite{klaers,schmitt2016}. In this system, both the occupation
statistics and first-order coherence properties of the condensate
have been experimentally
investigated \cite{schmitt2014,damm,schmitt2016_ssb,greveling2018},
providing evidence for the coexistence
of phase coherence and large number fluctuations of the condensate. 

This concomitant occurrence of macroscopic fluctuations and phase coherence found a consistent theoretical
framework in the recent paper~\cite{loro}, where it is shown
that the standard Bogoliubov quasi-average construction \cite{bogoliubov} fails
to reproduce the fluctuation properties observed in this setting,
which are known to be anomalously large \cite{ziff,weiss}.
All these observations motivate
an approach to the problem in which the condensate mode operator can be represented by a classical
phase and a fluctuating density operator.

In this paper we indeed show that, within the framework of the ideal gas, these results admit a simple formulation
in terms of a phase-density decomposition of the condensate mode \cite{lynch}.
Our analysis is based on two complementary arguments.
First, we show that in a phase-density representation the full hierarchy
of correlation functions is entirely determined by the statistics of the
condensate density. Second, since the condensate density follows an
exponential distribution in the grand canonical ensemble \cite{terHaar},
all correlation functions are obtained in closed form. 
This provides a transparent physical interpretation of photon
Bose-Einstein condensation in the grand canonical framework,
in which phase coherence and density fluctuations naturally coexist. 
This scenario is the dual counterpart of the familiar quasi-condensate regime
of low-dimensional interacting
atomic Bose gases, see for instance \cite{popov,petrov2000,petrov2000b} and the paper \cite{Mora}:
there, phase fluctuations dominate while density fluctuations are suppressed,
whereas here the phase is well defined and the amplitude fluctuates strongly.

In the last part of the paper, inspired by quantum optics techniques,
we propose a direct way to measure the square modulus of the anomalous
average of the condensate mode, which allows to experimentally assess
the theoretical predictions of the grand canonical approach.

\section{Phase-density representation of the Bose condensate}

We consider an ideal gas of non-interacting  bosons of mass $m$ described by
the second-quantized Hamiltonian 
\beq 
{\hat H} = \int d^3{\bf r} \ {\hat \psi}^{\dagger}({\bf r}) 
\left[ -\frac{\hbar^2}{2m} \nabla^2 + U({\bf r}) 
\right] {\hat \psi}({\bf r}),
\label{hamilton.0}
\eeq 
where ${\hat \psi}^{\dagger}({\bf r})$ and ${\hat \psi}({\bf r})$ are the
creation and annihilation operators of the bosonic matter field, respectively.
We first focus on the
$3d$ box of side $L$, with volume $V=L^3$, setting the confining potential
$U({\bf r}) \equiv 0$. The energy levels are then
\beq
  \epsilon_{\bf k} = \frac{\hbar^2}{2m}\, |{\bf k}|^2,
  \quad 
  k_j = \frac{2\pi}{L}\,   m_j,
  \quad m_j = 0,\pm 1,\pm 2, \ldots, 
\eeq
with $\quad j=x,y,z$.
The total number operator $\hat{N}$ is defined as
\beq 
\hat{N}=\sum_{\bf k}\hat{a}_{{\bf k}}^\dagger \hat{a}_{{\bf k}} = \sum_{\bf k}
    {\hat N}_{\bf k},
\eeq
 where
$\hat{a}^\dag_{{\bf k}}$ ($\hat{a}_{{\bf k}}$) is the creation
(annihilation) operator in the state $|{\bf k}\rangle$.
The matter field can be expressed in terms of the one-particle eigenfunctions
$\phi_{{\bf k}}({\bf r})=e^{i {\bf k}\cdot {\bf r}}/\sqrt{V}$,
\beq
\hat{\psi}({\bf r})=
\sum_{{\bf k}}\hat{a}_{{\bf k}}\phi_{{\bf k}}({\bf r})
\eeq
and, in particular,
\beq
\hat{\psi}_{{\bf 0}} = \frac{\hat{a}_{\bf 0}}{\sqrt{V}}.
\eeq
In the grand canonical ensemble (GCE) the average occupation number is given by
\beq
\langle {\hat N}_{\bf k}\rangle = \frac{1}{e^{\beta(\epsilon_{\bf k}-\mu)}-1},
\eeq
where the brackets stand for the thermal average
\beq
\langle \cdots \rangle = \frac{1}{{\cal Z}} \, \textrm{Tr}[e^{-\beta (\hat{H}-\mu \hat{N})} (\cdots)],
\label{av.1}
\eeq
with $\beta=1/(k_BT)$, $k_B$ the Boltzmann constant, $T$ the
temperature, $\mu$ the chemical potential and 
\beq 
{\cal Z}=\textrm{Tr}[e^{-\beta (\hat{H}-\mu \hat{N})}]
\eeq
the partition function.

\subsection{Bose-Einstein condensation and spontaneous symmetry breaking}

It is well known that in the thermodynamic limit $V\to \infty$, this system
can manifest Bose-Einstein condensation (BEC). Let us separate out of
the total average density $\rho=\sum_{\bf k}\langle {\hat N}_{\bf k}\rangle/V$ 
the ${\bf k}={\bf 0}$ contribution
\beq
\rho = \frac{1}{V}\frac{z}{1-z} + \rho^\prime,
\label{dens.1}
\eeq
where $z=e^{\beta \mu}$ is the fugacity and $\rho^\prime$ denotes the remaining contribution of the
excited states. Keeping the temperature constant and using $\rho$ as control parameter, there exists a critical
value $\rho_c$ above which in the thermodynamic limit $\rho^\prime$ saturates to 
$\rho_c$ and the
density of the ground state
$\rho_{\bf 0}=\langle {\hat N}_{\bf 0}\rangle/V$ acquires a finite value~\cite{ziff}
\beq
\rho_{\bf 0} =  \frac{1}{V}\frac{z}{1-z} \to \left \{ \begin{array}{ll}
         0, \;\; $for$ \;\; \Delta \rho \leq 0, \\
         \\
         \Delta \rho,\;\; $for$ \;\; \Delta \rho > 0,
        \end{array}
        \right .
        \label{BEC_3d.2}
        \eeq 
        where $\Delta \rho=\rho - \rho_c$. Notice that from the above relation there remains defined $z(\rho)$ as 
        a function of $\rho$~\cite{huang}.
        In this setting, the density fluctuations become macroscopic,
        producing the so-called grand
        canonical catastrophe  \cite{ziff,kruk,sala2024}
\beq
\lim_{L \to \infty}\frac{1}{V^2}
(\langle \hat{N}_{\bf 0}^2\rangle-\langle \hat{N}_{\bf 0}\rangle^2)  =
\left \{ \begin{array}{ll}
         0, \;\; $for$ \;\; \Delta \rho\leq 0, \\
         \\
         \Delta \rho^2,\;\; $for$ \;\; \Delta \rho > 0.
        \end{array}
        \right .
        \label{eqstate_3d.5}
        \eeq
              
These are well-known results, but some comments are in order.
The predictions on the fluctuations of the GCE have been considered
in the past literature as a pathology of this ensemble: macroscopic fluctuations
were supposed to be unphysical and therefore the GCE was regarded as flawed and not trustworthy
\cite{holthaus, fujiwara,grossman, weiss}. The prevalent attitude was then
to consider the   canonical ensemble as the correct framework,
where the issue of the large fluctuations does not occur.
According to another approach, the ``catastrophic''
fluctuations can be removed invoking the argument of spontaneous symmetry breaking
(SSB) \cite{yukalov2007}.

The occurrence of BEC is indeed related
to the spontaneous breaking of
the U(1) symmetry of the Hamiltonian~(\ref{hamilton.0}), which is invariant
under the (global) gauge transformation
$\hat{\psi}({\bf r})\to e^{i\theta}\hat{\psi}({\bf r})$ \cite{suto,lieb,WZ}.
From Eq.~(\ref{BEC_3d.2}) there follows that the average over the symmetric
state~(\ref{av.1}) yields~\cite{stringari}
\beq
\lim_{|{\bf r}-{\bf r}'|\to\infty} \langle \hat{\psi}^{\dagger}({\bf r})
\hat{\psi}({\bf r}')\rangle -\langle \hat{\psi}^{\dagger}({\bf r})
\rangle\langle \hat{\psi}({\bf r}') \rangle = \left \{ \begin{array}{ll}
         0, \;\; $for$ \;\; \Delta\rho \leq 0, \\
         \\
          \Delta\rho,\;\; $for$ \;\; \Delta \rho > 0, \\
          \end{array}
        \right .
        \label{eq.1}
\eeq 
namely for $\Delta\rho \leq 0$ the clustering property is satisfied implying that
$\langle \cdots\rangle$ is a pure state, while for
$\Delta\rho > 0$ the clustering property does not hold, long range order
develops in the system, revealing that 
the thermal state is a mixture of broken-symmetry pure states of the form
$\langle \cdots\rangle=\int d\theta/(2\pi)\langle \cdots\rangle_\theta$~\cite{ginibre,parisi}.
The problem then is how to build the proper  (and physically
meaningful) state $\langle \cdots\rangle_\theta$.
The standard approach relies on the Bogoliubov prescription \cite{bogo},
which requires the introduction in the Hamiltonian of a fictitious complex
external field, coupled to the zero mode operator, which is sent to zero only
after the thermodynamic limit is taken. This procedure defines the so-called
quasi-averages and is equivalent to the substitution of the zero-momentum matter
field operator with a $c$-number~\cite{ginibre}
\beq
\hat\psi_{\bf 0} \to \psi_{\bf 0} = e^{i\theta}\sqrt{\rho_{\bf 0}},
\label{assumption_bogo}
\eeq
which fixes both the phase and the amplitude, removing  
the grand canonical catastrophe.

However, this very same result shows that the
Bogoliubov procedure, in this particular case, fails to isolate the correct broken-symmetry state,
for the density operator $\hat{\rho}_{\bf 0}$ commutes with the generator of the U(1) symmetry. Therefore,
the density and its fluctuations must have the same expectation values in the symmetrical density 
matrix~(\ref{av.1}) and in each of the broken-symmetry states 
$\langle \cdots\rangle_\theta$ \cite{ginibre}, 
implying that the grand canonical catastrophe 
is not just a property of the mixed state, but that it holds at the basic level of
the pure states (see Appendix \ref{symmetric} for a detailed explanation). 
The experimental findings reported in
\cite{schmitt2014,schmitt2016_ssb}, showing that both the grand canonical catastrophe and SSB
are observed in the BEC of photons, are aligned with this prediction, as it should be.

On the theoretical side, the framework reconciling macroscopic fluctuations
and SSB has been recently proposed by some of us in Ref.~\cite{loro}.
Building on the Glauber-Sudarshan P representation of the density
matrix \cite{glauber,sudarshan}, the correct broken-symmetry
state which accounts for the large fluctuations of the density has been explicitly constructed.
In this new scenario the mechanism of condensation is quite different from the usual one operating
in the Bogoliubov picture of the quasi-averages. The core novelty is the replacement of the fixed
amplitude implied by the ansatz~(\ref{assumption_bogo}) by a fluctuating amplitude.
This is the key ingredient indispensable to keep together BEC, SSB and
the macroscopic fluctuations of the density. In the end all the pieces fall into place and
BEC in the GCE is recognized to be a transition driven by condensation of fluctuations \cite{zannetti,CSZ}.

\subsection{Phase-density representation of the bosonic field operator}

Motivated by the above considerations, here we present an alternative way to reproduce the
results presented in~\cite{loro}, resorting to a different ansatz for the field operator which fixes the
shortcomings of the Bogoliubov prescription, when applied in this particular context. 
The appeal of the new approach is of offering a 
more direct and intuitive understanding of the ways in which BEC in the CGE differs from
the standard picture. The key step is to assume that, in the broken-symmetry sector, 
the zero-momentum mode operator can be represented by the substitution
\beq
\hat\psi_{\bf 0} \to \hat{\Psi}_{\bf 0} \equiv e^{i\theta}\sqrt{\hat{\rho}_{\bf 0}},
\label{assumption}
\eeq
where $\theta$ is a classical phase and
${\hat\rho}_{\bf 0}=\hat{\psi}_{\bf 0}^\dagger\hat{\psi}_{\bf 0}=\hat{a}_{\bf 0}^\dagger \hat{a}_{\bf 0}/V$
is the condensate density operator.
Notice that this ansatz does not suffer from the inconsistency related to
the non-Hermitian nature of the phase operator  \cite{lynch}, because
here the phase is fixed. More importantly and at variance with the Bogoliubov ansatz (\ref{assumption_bogo}), here
the operatorial character of the amplitude is preserved, although by rendering it Hermitian.
It then follows immediately that
\beq
\hat\Psi^{\dagger n}_{\bf 0}\hat\Psi^m_{\bf 0}
= e^{-in\theta}\left(\hat{\rho}_{\bf 0}^{n/2}\right)^\dagger
e^{im\theta}\hat\rho_{\bf 0}^{m/2} =
e^{i(m-n)\theta}\,\hat\rho_{\bf 0}^{(n+m)/2},
\eeq
where we have used that $\hat{\rho}_{\bf 0}$ is positive Hermitian. 
Thus,
taking expectations on the thermal state, we obtain
\beq
\langle \hat\Psi^{\dagger n}_{\bf 0}\hat\Psi^m_{\bf 0}\rangle
=\int_0^{2\pi}\frac{d\theta}{2\pi} \langle \hat\Psi^{\dagger n}_{\bf 0}\hat\Psi^m_{\bf 0}\rangle_\theta,
\label{corr.1}
\eeq
where
\beq
\langle \hat\Psi^{\dagger n}_{\bf 0}\hat\Psi^m_{\bf 0}\rangle_\theta
=
e^{i(m-n)\theta}\,
\langle \hat\rho_{\bf 0}^{(n+m)/2}\rangle_\theta.
\label{corr.1bis}
\eeq
Since $\hat\rho_{\bf 0}^{(n+m)/2}$
is a symmetrical operator, as remarked above
the average $\langle \hat\rho_{\bf 0}^{(n+m)/2}\rangle_\theta$
can be computed on the thermal state finding
\beq
\langle \hat\Psi^{\dagger n}_{\bf 0}\hat\Psi^m_{\bf 0}\rangle_\theta
=
e^{i(m-n)\theta}\,
\langle \hat\rho_{\bf 0}^{(n+m)/2}\rangle, 
\label{corr.1}
\eeq
namely all correlation functions of the operator $\hat\Psi_{\bf 0}$ in the broken-symmetry state are determined by the moments
of the condensate density.
In the following we will show that the correlator (\ref{corr.1}) reproduces the
correct correlation functions of the zero-mode matter field operator $\hat{\psi}_{\bf 0}$ in the broken-symmetry state $\langle \hat\psi_{\bf 0}^{\dagger n}\hat\psi_{\bf 0}^m\rangle_\theta$, that have been computed exactly in Ref.~\cite{loro}
without any approximation.

Finally, let us note that the ansatz (\ref{assumption}) is somehow
complementary to the assumption usually introduced to treat
interacting atomic Bose gases in the very low temperature regime,
where density fluctuations are suppressed and a phase operator is
introduced to describe phase fluctuations, see for instance the
discussion in \cite{landau} and
Refs. \cite{popov,petrov2000,petrov2000b}.

\subsection{Extension to a generic confining potential}

The phase-density representation (\ref{assumption}) introduced above does not rely
on the specific geometry of a box, and can be extended to a generic
external potential $U({\bf r})$. 
In this case, it is natural to expand the bosonic field operator in the
basis $\phi_{{\boldsymbol \alpha}}({\bf r})$ of the single-particle eigenfunctions 
\beq
\hat\psi({\bf r})
=
{\hat a}_{\bf 0} \,\phi_{\bf 0}({\bf r})
+
\sum_{{\boldsymbol \alpha} \neq {\bf 0}} {\hat a}_{\boldsymbol \alpha} \,
\phi_{\boldsymbol \alpha}({\bf r}),
\eeq
where $\phi_{\bf 0}({\bf r})$ denotes the lowest-energy mode of the
confining potential, and $\hat a_{\bf 0}$ its annihilation operator.
Here ${\boldsymbol \alpha}$ labels the set of quantum numbers of the
single-particle
problem with single-particle energies $\epsilon_{\boldsymbol \alpha}$. 
In the presence of Bose-Einstein condensation, the mode $\phi_{\bf 0}({\bf r})$
is macroscopically occupied, while the contribution of the excited 
modes remains subleading in the thermodynamic limit.
The condensate field can therefore be identified with the operator
\beq
{\hat\psi}_{\bf 0}({\bf r}) = {\hat a}_{\bf 0} \,\phi_{\bf 0}({\bf r}).
\eeq
One can again use ${\hat N}_{\bf 0}={\hat a}_{\bf 0}^{\dagger}{\hat a}_{\bf 0}$
and introduce the ansatz
\beq
    {\hat\psi}_{\bf 0}({\bf r}) \to \hat\Psi_{\bf 0}({\bf r}) \equiv e^{i\theta} \, \sqrt{{\hat N}_{\bf 0}} \,
    \phi_{\bf 0}({\bf r}),
\eeq
so that the full hierarchy of correlation functions of the condensate
mode is determined by the moments of the number operator $\hat{N}_{\bf 0}$,
exactly as in the homogeneous case. 
This shows that the phase-density formulation is not tied to
translational invariance, but applies to any system in which a single
mode is macroscopically occupied. The spatial structure of the
condensate enters only through the wavefunction $\phi_{\bf 0}({\bf r})$,
while its statistical properties are encoded in the operator
${\hat N}_{\bf 0}$.

\subsection{Distribution of the condensate density in the grand-canonical
  ensemble}
  
Let us now go back to the central result~(\ref{corr.1}) and to the evaluation of the right hand side.
Using the spectral representation of the operator
\beq
\hat{\rho}_{\bf 0}^q = \sum_{N_{\bf 0}=0}^\infty \left( \frac{N_{\bf 0}}{V} \right)^q | N_{\bf 0} \rangle \langle N_{\bf 0}|,
\label{corr.2}
\eeq
with $|N_{\bf 0}\rangle$ the Fock state such that
${\hat N_{\bf 0}}|N_{\bf 0}\rangle =
N_{\bf 0} |N_{\bf 0}\rangle$, the thermal expectation is given by
\beq
\langle \hat{\rho}_{\bf 0} ^q \rangle = \sum_{N_{\bf 0}=0}^\infty P(N_{\bf 0}) \left( \frac{N_{\bf 0}}{V} \right)^q,
\label{corr.3}
\eeq
where the probability distribution to have $N_{\bf 0}$ particles in the ground state is given by~\cite{ziff}
\beq
P(N_{\bf 0}) = (1 - z)\, z^{N_{\bf 0}}.
\label{distr1}
\eeq
Now, recalling that the total average density $\rho$ is the control parameter, 
the function $z(\rho)$ must be inserted into the right hand side. Using Eq.~(\ref{BEC_3d.2}), one has $z=V\Delta\rho/(1+V\Delta\rho)$, where the
probability of $N_{\bf 0}$, conditioned to the chosen value of $\rho$, in the condensed phase for large $V$
is given by 
\beq
P(N_{\bf 0}|\rho) \sim \frac{1}{V\Delta\rho}
e^{-N_{\bf 0}/(V\Delta\rho)}.
\label{PN0}
\eeq
Inserting into Eq.~(\ref{corr.3}) and transforming the sum into an integral we get
\beq
\langle \hat{\rho}_{\bf 0} ^q \rangle = \int_0^\infty dx \, p(x|\rho) \, x^q,
\label{corr.4}
\eeq
with
\beq
p(x|\rho) = \frac{1}{\Delta\rho} \, e^{-x/\Delta\rho}.
\label{corr.5}
\eeq
Carrying out the integration, one finds
\beq
\langle {\hat\rho}_{\bf 0}^q\rangle
=
\Gamma(q+1)\,\Delta\rho^q,
\eeq
where $\Gamma(x)$ is the Euler gamma function.
Substituting into the phase-density representation (\ref{corr.1}), this yields
\beq
\langle\hat\Psi^{\dagger n}_{\bf 0} \hat\Psi^m_{\bf 0} \rangle_\theta=
\Gamma\!\left(\frac{n+m}{2}+1\right)
e^{i(m-n)\theta}
\Delta\rho^{(n+m)/2},
\label{result}
\eeq
which indeed reproduces the exact result in Eq.~(18) of Ref.~\cite{loro} for the quantity $\langle\hat\psi_{\bf 0}^{\dagger n} \hat\psi_{\bf 0}^m \rangle_\theta$. It must be stressed
that the above result has been obtained by computing directly the expectation in the thermal state,
whereas the computation in~\cite{loro} did require to extract
the pure state $\langle \cdots \rangle_\theta$ from the mixed state. Therefore,
the approximation involved in the ansatz~(\ref{assumption}) is good enough to preserve the exact
values of all the moments of the field operators.

In particular, taking $n=0$ and $m=1$ one gets 
\beq
\langle \hat{\Psi}_{\bf 0} \rangle_\theta =
\Gamma(3/2)\,e^{i\theta}\sqrt{\Delta\rho},
\label{result1}
\eeq
which reproduces Eq.~(20)
of Ref.~\cite{loro} for the anomalous average $\langle \hat\psi_{\bf 0} \rangle_\theta$.
From this result important consequences follow. The first one is that, since $\Gamma(3/2)^2
=\pi/4$, the density is strictly greater
than the amplitude square of the anomalous average
\beq
\langle\hat\Psi^{\dagger}_{\bf 0} \hat\Psi_{\bf 0} \rangle_\theta >
|\langle \hat\Psi_{\bf 0} \rangle_\theta|^2.
\label{ampl.1}
\eeq
Therefore, here we have a significant departure from the standard Bogoliubov picture,
since the formation of a finite anomalous average with a fixed phase is not enough to account for
the condensate density. Rather, there is a macroscopically-large 
difference which must be
made up by the fluctuations of the field amplitude
\beq
\langle \hat{\rho}_{\bf 0} \rangle = \langle\hat\Psi^{\dagger}_{\bf 0} \hat\Psi_{\bf 0} \rangle_\theta =
|\langle \hat\Psi_{\bf 0} \rangle_\theta|^2 + 
\langle \delta \hat{\Psi}_{\bf 0}^\dagger  \delta \hat{\Psi}_{\bf 0}\rangle_\theta
\label{ampl.2}
\eeq
where $\delta \hat{\Psi}_{\bf 0} = \hat{\Psi}_{\bf 0} - \langle \hat{\Psi}_{\bf 0} \rangle_\theta$ and
\beq
\langle \delta \hat{\Psi}_{\bf 0}^\dagger  \delta \hat{\Psi}_{\bf 0}\rangle_\theta =
[1- \Gamma(3/2)^2] \, \Delta\rho \;.
\eeq
As discussed in detail in \cite{loro}, this result reveals the different
nature of BEC under GCE conditions, as an instance of condensation of
fluctuations. 

The second remarkable consequence is that the grand canonical catastrophe can now be
recognized to be nothing but the observable manifestation of these underlying fluctuations 
at the higher level of the density fluctuations. 
In fact, working out the algebra, the result of the second line of Eq.~(\ref{eqstate_3d.5})
is borne out of the existence of $\delta \hat{\Psi}_{\bf 0}$ via the relation
\beqa
 \langle(\hat\Psi^{\dagger}_{\bf 0} \hat\Psi_{\bf 0})^2 \rangle_\theta - 
\langle \hat\Psi^{\dagger}_{\bf 0} \hat\Psi_{\bf 0} \rangle_\theta^2 &=&
4|\langle \hat\Psi^{\dagger}_{\bf 0}\rangle_\theta|^2  \langle \delta \hat\Psi^{\dagger}_{\bf 0} \delta \hat\Psi_{\bf 0}\rangle_\theta 
 \nonumber \\
&+ &  \langle (\delta \hat\Psi^{\dagger}_{\bf 0} \delta \hat\Psi_{\bf 0})^2\rangle_\theta
- \langle \delta \hat\Psi^{\dagger}_{\bf 0} \delta \hat\Psi_{\bf 0}\rangle_\theta^2 \nonumber \\
&+& 4|\langle \hat\Psi^{\dagger}_{\bf 0}\rangle_\theta| 
\langle |\delta \hat\Psi_{\bf 0}|^3\rangle_\theta \nonumber \\
& = & \Delta\rho^2.
\label{GCC}
\eeqa
Clearly, if $\delta \hat\Psi_{\bf 0}$ vanishes, as in the Bogoliubov procedure, there is no
grand canonical catastrophe (see Sect. \ref{symmetric}).

Summarising, we have the following picture of BEC in the GCE: by fixing the total average density $\rho$
to a value above the critical one $\rho_c$, Eq.~(\ref{dens.1}) acts as a constraint which forces the condensate
density to acquire the finite value $\Delta \rho$. However, this cannot be achieved simply by breaking
the symmetry and by meeting the $\Delta \rho$ requirement by developing a sufficiently large anomalous average.
In the GCE this doesn't work, because the anomalous average falls short of $\Delta \rho$ and, therefore,
there is a gap which is filled up by letting the matter field $ \hat\Psi_{\bf 0}$ to undergo fluctuations of the
size appropriate to balance the density. The additional motive of interest, and not a minor one, 
is that these fluctuations propagate
to the mesoscopic level of the density fluctuations through Eq.~(\ref{GCC}) 
accounting for the controversial grand canonical catastrophe.

In Appendix \ref{fock} we discuss how the same results can be obtained
enforcing the ansatz (\ref{assumption}) in the broken-symmetry states
in the Fock representation.

\section{Experimental signatures and observables}

In the photon-condensate setting mentioned in the Introduction,
Bose-Einstein condensation in the grand-canonical ensemble is characterized
by the coexistence of a classical phase $\theta$ (phase coherence)
and a strongly fluctuating condensate density, whose moments grow factorially. 
The failure of the $c$-number substitution is directly traced to the
nontrivial statistics of $\hat\rho_{\bf 0}$, while phase coherence is preserved
through the factor $e^{i(m-n)\theta}$. The condensate can thus be
viewed as a coherent field with a strongly fluctuating amplitude. 
In the presence of a confining potential, the condensate corresponds
to a spatial mode $\phi_{\bf 0}({\bf r})$, and the associated operator
$\hat a_{\bf 0}$ can be accessed experimentally by projecting the field
$\hat\psi({\bf r})$ onto this mode. In optical implementations, such
as photon Bose-Einstein condensates, this projection is naturally
realized by mode-selective detection \cite{damm,greveling2018}.

A direct observable consequence of the phase-density structure is
obtained by comparing the coherent field amplitude with the condensate
density. In the broken-symmetry sector we found the result in Eq. (\ref{result1}) and,
from Eq. (\ref{result}) with $n=m=1$, 
$\langle \hat\psi_{\bf 0}^\dagger \hat\psi_{\bf 0} \rangle = \rho_{\bf 0}$ 
(we omit the subscript $\theta$ from here on and return to the condensate mode operator $\psi_{\bf 0}$ to lighten the notation),
so that the ratio between coherent amplitude and condensate density is
\beq
\mathcal R
\equiv
\frac{|\langle \hat\psi_{\bf 0} \rangle|^2}
     {\langle \hat\psi_{\bf 0}^\dagger \hat\psi_{\bf 0} \rangle}
=
\Gamma(3/2)^2
=
\frac{\pi}{4}.
\eeq
This result should be contrasted with the usual $c$-number description
of a condensate mode, which would imply $\mathcal R = 1$. 
The present framework therefore predicts
a characteristic reduction of the coherent
fraction of the condensate mode: only a fraction $\pi/4$ of the
condensate density contributes coherently to the field amplitude.
The ratio $\mathcal R$ thus provides a direct and experimentally
accessible signature of the interplay between coherence and number
fluctuations. However, it should be also stressed that, in real experiments, other possible effects could contribute to a reduction of the parameter $R$, such as thermal fluctuations at finite temperature, spontaneous losses and shot noise.

It is instructive to compare this behavior with that of laser light.
A single-mode laser field above threshold is well described by a
coherent state, characterized by a well-defined phase and Poissonian
photon-number fluctuations \cite{glauber,sudarshan,mandel}. In this
ideal case one has
$|\langle \hat a \rangle|^2 = \langle \hat a^\dagger \hat a \rangle$,
so that the coherent fraction is equal to unity.
Deviations from this behavior can occur in the presence of phase
diffusion or classical fluctuations of the field amplitude, where the
quantum state is described by a statistical mixture of coherent states
\cite{mandel,scully}. In such situations one generally finds
$|\langle \hat a \rangle|^2 < \langle \hat a^\dagger \hat a \rangle$.
In contrast, the present framework predicts a reduced coherent fraction
$\mathcal R = \pi/4$ even in the presence of a defined phase.
This shows that the suppression of $\mathcal R$ is not necessarily
associated with phase decoherence, but can arise from intrinsic,
strongly non-Poissonian amplitude fluctuations of the condensate mode.

In experiments with condensed photons,
the ratio $\mathcal R$ can be accessed by combining
measurements of the condensate intensity,
$\langle \hat a_{\bf 0}^\dagger \hat a_{\bf 0} \rangle$, with phase-sensitive
measurements of the field amplitude $\langle \hat a_{\bf 0} \rangle$, using
for instance homodyne or heterodyne detection \cite{mandel,scully}. 
Such measurements appear particularly natural in photon
Bose-Einstein condensates, where both intensity fluctuations and
first-order coherence have already been probed \cite{damm,greveling2018}.
To our knowledge, no direct measurement of $\mathcal{R}$ has been
reported so far in photon BEC experiments, making this ratio an open
and accessible experimental challenge.
In this context, $\mathcal R$ provides a direct and quantitative test
of a condensate with a fixed phase but a strongly fluctuating amplitude. 

More generally, the deviation of $\mathcal R$ from unity reflects the
fact that, see Eq.~(\ref{ampl.1}),
\beq
|\langle \hat\psi_{\bf 0} \rangle|^2 \neq
\langle \hat\psi_{\bf 0}^\dagger \hat\psi_{\bf 0} \rangle,
\eeq
which is the hallmark of a condensate that is not described by a
single coherent state, but rather by a superposition with a
nontrivial amplitude distribution. The ratio
${\mathcal R}= |\langle \hat a_{\bf 0}\rangle|^2/
\langle \hat a_{\bf 0}^\dagger \hat a_{\bf 0}\rangle$
is closely related to first-order coherence, since it compares the squared
mean field amplitude with the mode intensity. Unlike the usual second-order
correlator
\beq 
g^{(2)}(0)=\frac{\langle \hat a_{\bf 0}^\dagger \hat a_{\bf 0}^\dagger
  \hat a_{\bf 0} \hat a_{\bf 0}\rangle}
{\langle \hat a_{\bf 0}^\dagger \hat a_{\bf 0}\rangle^2},
\eeq
however, $\mathcal R$ is not determined by the second moment alone, but depends
on the full distribution of the condensate amplitude. For the exponential
statistics considered here one finds simultaneously
\beq
g^{(2)}(0) = 2
\qquad \text{and} \qquad
\mathcal R = \frac{\pi}{4},
\eeq
showing that strong bunching and a reduced coherent fraction can coexist
with a well-defined condensate phase. The prediction $g^{(2)}(0) = 2$
corresponds to the ideal grand canonical limit of an infinite reservoir.
In the experiments of Schmitt et al.~\cite{schmitt2014,schmitt2016_ssb}, values of
$g^{(2)}(0)$ up to $1.93$ have been measured in the condensed phase,
interpolating between the canonical value $g^{(2)}(0) = 1$ and the
grand-canonical limit $g^{(2)}(0) = 2$ as the reservoir size is increased.
This behavior is fully consistent with the present framework, and suggests
that the coherent fraction $\mathcal{R}$ should similarly interpolate
between $1$ (canonical limit) and $\pi/4$ (grand-canonical limit) as a
function of reservoir size --- a dependence that could be directly probed
in existing experimental setups.
While $g^{(2)}(0) \simeq 2$ has thus been confirmed experimentally
\cite{schmitt2014}, the value $\mathcal{R} = \pi/4$ constitutes a 
distinct and independent prediction that has not yet been tested.

\subsection{Interferometric measurement of the coherent fraction}

Inspired by standard quantum-optics techniques \cite{mandel,scully}, we
suggest that the coherent fraction ${\cal R}$ can be accessed by
interfering the condensate mode $\hat a_{\bf 0}$ with an external coherent
reference field. Let the reference field be described by a coherent
amplitude
\beq
\alpha = |\alpha|e^{i\phi},
\eeq
where $\phi$ is a controllable phase. The condensate mode and the
reference field are mixed on a balanced beam splitter. The field operator
at one output port is then
\beq
\hat b =
\frac{\hat a_{\bf 0}+\alpha}{\sqrt{2}},
\eeq
where the reference field has been treated as a classical coherent
amplitude.

The average intensity at this output port is
\beqa
I(\phi)
&=&
\langle \hat b^\dagger \hat b\rangle
\nonumber \\
&=&
\frac{1}{2}
\left[
\langle \hat a_{\bf 0}^\dagger \hat a_{\bf 0}\rangle
+
|\alpha|^2
+
2\,\mathrm{Re}\!\left(\alpha^*\langle \hat a_{\bf 0}\rangle\right)
\right].
\eeqa
Writing
\beq
\langle \hat a_{\bf 0}\rangle =
|\langle \hat a_{\bf 0}\rangle|e^{i\theta},
\eeq
one obtains
\beq
I(\phi)
=
\frac{1}{2}
\left[
\langle \hat N_{\bf 0}\rangle
+
|\alpha|^2
+
2|\alpha|\,|\langle \hat a_{\bf 0}\rangle|
\cos(\phi-\theta)
\right],
\eeq
with $\langle \hat N_{\bf 0}\rangle =
\langle \hat a_{\bf 0}^\dagger \hat a_{\bf 0}\rangle$.

By varying the phase $\phi$, the extremal intensities are
\begin{align}
I_{\max}
&=
\frac{1}{2}
\left[
\langle \hat N_{\bf 0}\rangle
+
|\alpha|^2
+
2|\alpha|\,|\langle \hat a_{\bf 0}\rangle|
\right],
\\
I_{\min}
&=
\frac{1}{2}
\left[
\langle \hat N_{\bf 0}\rangle
+
|\alpha|^2
-
2|\alpha|\,|\langle \hat a_{\bf 0}\rangle|
\right].
\end{align}
Therefore,
\beq
|\langle \hat a_{\bf 0}\rangle|
=
\frac{I_{\max}-I_{\min}}{2|\alpha|}.
\eeq
The condensate population $\langle \hat N_{\bf 0}\rangle$ can be obtained from
an independent intensity measurement of the condensate mode in the absence
of the reference field. Combining the two measurements gives
\beq
{\cal R}
=
\frac{|\langle \hat a_{\bf 0}\rangle|^2}
{\langle \hat a_{\bf 0}^\dagger \hat a_{\bf 0}\rangle}
=
\frac{(I_{\max}-I_{\min})^2}
{4|\alpha|^2\langle \hat N_{\bf 0}\rangle}.
\eeq
This interferometric protocol directly measures the coherent field
amplitude $\langle \hat a_{\bf 0}\rangle$, which cannot be extracted from
intensity measurements alone, and therefore provides access to the
coherent fraction ${\cal R}$.

\section{Conclusion}

We have shown that the properties of photon Bose-Einstein condensation in the 
grand canonical ensemble can be reformulated within a phase-density
representation where the phase is classical while the condensate
amplitude remains strongly fluctuating. In this framework, the full
hierarchy of correlation functions follows from 
the statistics of the condensate density. 
Within the phase-density framework considered here, where the phase is
assumed to be well defined while the condensate density fluctuates,
macroscopic occupation of a single mode does not imply a coherent state:
even in the limit of unit condensate fraction, the coherent fraction
remains smaller than one, with the explicit value $\mathcal{R} = \pi/4$
in the grand-canonical ensemble.

In real systems, this theoretical picture is expected to hold close to the transition threshold,
where interactions are negligible and the particle reservoir remains large enough to make
the grand canonical description effective \cite{schmitt2014}. Moreover, the phase coherence is expected to
persist over a characteristic time that depends on the size of the system \cite{schmitt2016_ssb}.   

The scenario presented here is dual to the quasi-condensate regime of
low-dimensional interacting
atomic Bose gases \cite{popov,petrov2000,petrov2000b},
where phase fluctuations dominate and density fluctuations are suppressed.
The two regimes thus represent opposite limits of the same general
decomposition: phase-stable/amplitude-fluctuating vs.\
amplitude-stable/phase-fluctuating.
Our results highlight a clear distinction between condensation and coherence,
and identify $\mathcal{R}$ as a direct experimental observable that
distinguishes the grand canonical condensate from both a coherent state
($\mathcal{R} = 1$) and a fully incoherent one ($\mathcal{R} = 0$).
A measurement of $\mathcal{R}$ via homodyne or heterodyne detection of
the condensate mode, combined with independent intensity measurements,
would provide a definitive test of this picture in current photon
BEC experiments. 

\section*{Acknowledgements}
AS acknowledges useful discussions with Prof. I. Carusotto and Prof. A. Rubano.
LS is partially supported by the `Iniziativa Specifica Quantum' of
INFN, by the Project ``Frontiere Quantistiche'' (Dipartimenti di Eccellenza)
of the Italian Ministry of University and Research
(MUR), by the European Union-Next Generation EU within the National
Center for HPC, Big Data and Quantum Computing
(Project No. CN00000013, CN1 Spoke 10: `Quantum Computing'),
and by the EU Project PASQuanS 2
`Programmable Atomic Large-Scale Quantum Simulation'.
AS acknowledges funding from the Italian Ministero dell'Universita'
e della Ricerca under the programme PRIN2022
("re-ranking of the final lists"), number2022KWTEB7, cup B53C24006470006.

\appendix
\section{Average of symmetrical operators}
\label{symmetric}

Since pure states in the condensed phase are identified by a fixed phase $\theta$, defining $\langle\cdots \rangle_\theta$ as the average in the broken-symmetry state,
we have that the symmetrical state (mixture of pure states) is represented as $\langle\cdots \rangle = \int d\theta/(2\pi) \langle\cdots \rangle_\theta$. Then, if we consider a symmetrical operator 
$\hat{A}$ that commutes with the gauge operator $\hat{U}(\theta)=e^{i\hat{N}\theta}$ which rotates the state of an angle $\theta$, we get
\beq
\langle\hat{A} \rangle_\theta = \langle  \hat{U}(\theta)^{\dagger} \hat{A} \hat{U}(\theta)\rangle_0 = \langle \hat{A} \rangle_0,
\label{A1}
\eeq
namely the average is independent of $\theta$, and therefore, using Eq. (\ref{A1}),
\beq
\langle\hat{A} \rangle = \int_0^{2\pi} \frac{d\theta}{2\pi} \langle\hat{A} \rangle_\theta = \langle\hat{A} \rangle_0.
\eeq 
This result implies that the Bogoliubov prescription (\ref{assumption_bogo}) cannot reproduce the correct broken-symmetry state. Indeed, focusing for instance on the density fluctuations and choosing $\hat{A}=(\hat{\psi}_{\bf 0}^{\dagger}\hat{\psi}_{\bf 0}-\rho_{\bf 0})^2$, from Eq. (\ref{eqstate_3d.5}) one gets $\langle\hat{A}\rangle=\Delta\rho^2$, whereas, using (\ref{assumption_bogo}) to construct $\langle\cdots \rangle_{\theta}$, one would get $\langle\hat{A} \rangle_{\theta}=0$,
leading to the inconsistency 
\beq
\langle\hat{A} \rangle \neq \langle\hat{A} \rangle_{\theta}.
\eeq 

\section{Broken-symmetry states in the Fock representaion}
\label{fock}

The Bogoliubov quasi-average (qa)
prescription, which is equivalent to the
ansatz~(\ref{assumption_bogo}), consists in the introduction of
a fictitious external field in the Hamiltonian, which is coupled with the zero mode operator
and explicitly breaks the gauge symmetry (see for instance \cite{ginibre}). Sending to zero this field after the thermodynamic limit, one gets
a broken-symmetry density
matrix in the form
\be
\hat{D}^{\textrm{(qa)}}_\theta=|e^{i\theta}\sqrt{\Delta\rho}\rangle
\langle e^{i\theta}\sqrt{\Delta\rho}|,
\eeq
that corresponds to
the projector onto the coherent state $|r e^{i\theta}\rangle$, which is eigenvector of $\hat{\psi}_{\bf 0}=\hat{a}_{\bf 0}/\sqrt{V}$ with eigevalue $r e^{i\theta}$,
where the amplitude $r=\sqrt{\Delta\rho}$ is the square root of the condensate density, and the phase
$\theta$ is selected by the vanishing external field.

On the other hand, as discussed in detail in Ref.~\cite{loro}, when the regular average (ra) prescription is adopted, namely when the external field is sent to zero before the thermodynamic limit, the broken-symmetry density matrix
takes the form
\beq
\hat{D}^{\textrm{(ra)}}_\theta=2\int_0^\infty dr~r \frac{1}{\Delta\rho}e^{-r^2/\Delta\rho}|e^{i\theta}r\rangle
\langle e^{i\theta}r|,
\eeq
which represents a weighted superposition of coherent states with different amplitudes and fixed phase.
Here we discuss how these opeartors can be written in terms of Fock
states. This is useful to clarify how
to define a phase-selected many-body quantum state
$|\theta\rangle$ which reproduces the  averages corresponding
to the different
assumptions~(\ref{assumption_bogo}) and~(\ref{assumption}). 

Starting from the relation between coherent states $\ket{\alpha}$ and Fock states $\ket{n}$~\cite{glauber}
\be
|\alpha\rangle = e^{-|\alpha|^2/2}\sum_{n=0}^{\infty}\frac{\alpha^n}{\sqrt{n!}}|n\rangle,
\ee
where $\hat{a}_{\bf 0}\ket{\alpha}=\alpha\ket{\alpha}$ and $\hat{a}^\dagger_{\bf 0}\hat{a}_{\bf 0}\ket{n}=n\ket{n}$, in the case of the quasi-averages one immediately gets
\be
\hat{D}^{\textrm{(qa)}}_\theta=e^{-\Delta N}\sum_{n,m=0}^{\infty}e^{i\theta(n-m)}\frac{\Delta N^{(n+m)/2}}{\sqrt{n!}\sqrt{m!}}\ket{n}\bra{m},
\ee
with $\Delta N=V\Delta\rho $.
Since the dependence on $n$ and $m$ factorizes, this expression can be rewritten in the form
\be
\hat{D}^{\textrm{(qa)}}_\theta=|\theta\rangle \langle\theta |
\ee
where
\be
|\theta\rangle=\sum_{n=0}^\infty e^{i\theta n}\sqrt{P_{\textrm{qa}}(n)}\ket{n},
\label{theta_bogo}
\ee
with
\be
   P_{\textrm{qa}}(n)  = \E^{-\Delta N} \frac{\Delta N^n}{n!},
   \quad\text{(Poisson distribution)}
\ee
the occupation number distribution in the Bogoliubov quasi-average state.
   Then, it is easy to compute
\be
  \ave{\hat{\psi}_{\bf 0}^{\dag n} \hat{\psi}_{\bf 0}^m}^{\textrm{(qa)}}_\theta = \langle\theta| \hat{\psi}_{\bf 0}^{\dag n} \hat{\psi}_{\bf 0}^m |\theta\rangle = \E^{i (n-m)\theta} \Delta \rho^{(n+m)/2},
\ee
where $\ave{\cdots}^{\textrm{(qa)}}_\theta = \textrm{Tr}[\hat{D}^{\textrm{(qa)}}_\theta \cdots ]$. This result
coincides with the one obtained with the Bogoliubov ansatz (\ref{assumption_bogo}).

On the contrary, in the case of the regular averages we have
\beqa
\hat{D}^{\textrm{(ra)}}_\theta &=&
2\int_0^{\infty}dr~r \frac{e^{-r^2/\Delta N}}{\Delta N}\left(e^{-r^2/2} \sum_{n=0}^\infty \frac{e^{i\theta n}r^n}{\sqrt{n!}}|n\rangle\right) \nonumber \\
&\times& \left(e^{-r^2/2} \sum_{m=0}^\infty \frac{e^{-i\theta m}r^m}{\sqrt{m!}}\langle m|\right)\nonumber \\
&=& \frac{2}{\Delta N} \sum_{n,m=0}^\infty \frac{e^{i\theta(n-m)}}{\sqrt{n!}\sqrt{m!}} \nonumber \\
&\times& \int_0^{\infty}dr~r^{1+n+m} e^{-r^2(1+1/\Delta N)}\ket{n}\bra{m}  \nonumber \\
%&=& \sum_{n,m=0}^\infty \frac{e^{i\theta(n-m)}}{\sqrt{n!}\sqrt{m!}}  \Delta N^{(n+m)/2}(1+\Delta N)^{-(2+n+m)/2}\Gamma[(2+n+m)/2]\ket{n}\bra{m}  \nonumber \\
%&=& \frac{1}{1+\Delta N}\sum_{n,m=0}^\infty \frac{e^{i\theta(n-m)}}{\sqrt{n!}\sqrt{m!}}  \left(\frac{\Delta N}{1+\Delta N}\right)^{(n+m)/2}\Gamma[(2+n+m)/2]\ket{n}\bra{m} \nonumber \\
&=& \frac{1}{1+\Delta N} \times \nonumber \\
 \sum_{n,m=0}^\infty  & & e^{i\theta(n-m)} \left(\frac{\Delta N}{1+\Delta N}\right)^{(n+m)/2}\frac{\left(\frac{n}{2}+\frac{m}{2}\right)!}{\sqrt{n!}\sqrt{m!}}\ket{n}\bra{m}. \nonumber \\
\eeqa
Since the factorial under the sum appearing in the last expression cannot be factorized in the product of functions of $n$ and $m$, in general a state $|\theta\rangle$ analogous to (\ref{theta_bogo}) cannot be defined. 
However, if we are interested in reproducing the result (\ref{result}), namely we restrict to consider averages of operators with the structure of Eq. (\ref{assumption}), then it is possible to define a fixed-phase state that in this case reads
\be
\label{eq:state_t_cs} 
  \ket{\theta} = \sum_{n=0}^\infty \E^{i n\theta}\, \sqrt{P_{\textrm{ra}}(n)} \ket{n},
\ee
where 
\be
   P_{\textrm{ra}}(n) = \frac{1}{1+\Delta N}\left(\frac{\Delta N}{1+\Delta N}\right)^{n},
\ee
such that
 \beqa
    \ave{\hat{\Psi}_{\bf 0}^{\dag n} \hat{\Psi}^m_{\bf 0}}_\theta^{\textrm{(ra)}}  &=&  \bra{\theta}  \hat{\Psi}_{\bf 0}^{\dag n} \hat{\Psi}_{\bf 0}^m\ket{\theta} \nonumber \\
    &=&\sum_{k,l}\bra{k}e^{-ik\theta}\sqrt{P_{\textrm{ra}}(k)} e^{-i\theta n}\left(\hat{\rho}_{\bf 0}^{n/2}\right)^\dag \nonumber \\
    &\times& e^{i\theta m}\hat{\rho}_{\bf 0}^{m/2}\sqrt{P_{\textrm{ra}}(l)}e^{il\theta}\ket{l} \nonumber \\
    &=&e^{i(m-n)\theta}\sum_k \left(\frac{k}{V}\right)^{(n+m)/2} P_{\textrm{ra}}(k), \label{Pra} \nonumber \\
%    &=& e^{i(m-n)\theta}\Gamma\left(\frac{n+m}{2}+1\right) \, \Delta\rho^{(n+m)/2}, \nonumber \\
\eeqa
where $\ave{\cdots}^{\textrm{(ra)}}_\theta = \textrm{Tr}[\hat{D}^{\textrm{(ra)}}_\theta \cdots ]$.
Now, taking the thermodynamic limit,
\beq
  P_{\textrm{ra}}(n) = \frac{1}{1+\Delta N} \left(1+ 1/\Delta N\right)^{-n}
             \underset{\Delta N \gg 1} {\sim}
                \frac{1}{\Delta N}\, \E^{- n/\Delta N},
\eeq
and, substituting the sum with an integral in Eq. (\ref{Pra}), one gets 
\beq
  \ave{\hat{\Psi}_{\bf 0}^{\dag n} \hat{\Psi}_{\bf 0}^m}^{\textrm{(ra)}}_\theta =
  \E^{i (n-m)\theta}\Gamma\left(1+\frac{n+m}{2}\right) \Delta \rho^{(n+m)/2}, 
\eeq
which is indeed in agreement with Eq.~(\ref{result}).
This shows that the state $|\theta\rangle$ in Eq. (\ref{eq:state_t_cs}) displays a well-defined phase $\theta$ together with
non-Poissonian number fluctuations, and,
within the phase-density representation~(\ref{assumption}),
reproduces the correct correlation functions.

\end{document}